# Optical and magnetic properties of Co-TiO$_2$ sandwich Composite films grown by magnetron sputtering

Fa-Min Liu [1, 2], Tian-Min Wang [1], J. Q. Li [2], Y. Q. Zhou [2], M. C. Zhang [3]

[1] Center of Material Physics and Chemistry, School of Science, Beijing University of Aeronautics and Astronautics, Beijing 100083, P. R. China

< Tel. +86-10-82317941, Fax. +86-10-82315933, E-mail: fmliu@etang.com ; fmliu@aphy.iphy.ac.cn >

[2] National Laboratory for Superconductivity, Institute of Physics, Chinese Academy of Sciences, Beijing 100080, P. R. China

[3] State Key Laboratory for Advanced Metals and Materials, University of Science and Technology Beijing, Beijing 100083, P. R. China

Abstract:

   The transition metal ions doped with TiO$_2$ films are of magnetic properties at room temperatures and have stimulated many scientists to study other properties. The Co-TiO$_2$ sandwich composite films have been recently grown on glass and silicon substrates by using alternately radio frequency reactive and direct current magnetron sputtering. The microstructural properties of these films were characterized with Raman spectra and X-ray photoemission spectra (XPS). It shows an anatase TiO$_2$ containing Co nano-layer. Optical absorption spectra have red shift with Co nano-layer increasing, which originated from the quantum confinement and tunnel effects. Magnetic properties show that the saturation magnetization, remanent magnetic induction and coercivity are about 4.35 emu /g, 1.02 emu/g and 569.6 Oe respectively for 12 nm Co layer between the two TiO$_2$ layers.

1. Introduction

Titanium dioxide (TiO$_2$) is an attractive material for application such as photocatalist [1-4], gas sensor [5] and gate oxides in MOS transistors [6]. Co-doped TiO$_2$ anatase, grown by pulsed laser deposition (PLD), has very recently been demonstrated to be weakly ferromagnetic (FM) and semiconducting for doping levels up to ~8 at. %, and temperatures of up to 400 K [7]. The magnetic properties of oxygen-plasma-assisted molecular-beam epitaxy (OPA-MBE) grown material [8] are significantly better than those of material grown analogous pulsed laser deposition. However, there is a few works dealing with optical and magnetic properties of Co-TiO$_2$ sandwich composite films. Recently, we have been able to grow Co-anatase TiO$_2$ sandwich multi-layer films on glass and silicon substrates by using alternately radio-frequency reactive (RFR) and direct current (DC) magnetron sputtering. In this paper, we report microstructure, optical and magnetic properties of the films, studied by X-ray diffraction, X-ray photoelectron spectroscopy, Raman spectroscopy, UV-visible spectrophotometer and vibrating sample magnetometer. We discuss the influence of Co nano-layer and substrate temperature on the optical characteristics of TiO$_2$ films based on optical parameters such as optical band gap and refractive index.

2. Experimental procedure

The Co-TiO$_2$ sandwich composite films deposited by alternately using radio frequency reactive and direct current magnetron sputtering. Glass and silicon substrates (20 mm × 20 mm) were ultrasonically washed successively in acetone, alcohol, and deionized water to obtain a clean surface before being placed in a vacuum chamber. And the substrate was further inversely sputtered for 5 minutes before the high film deposited. The target was separated from the substrate by 5 cm. First, high anatase $TiO_2$ was deposited by radio frequency reactive magnetron sputtering. When the chamber was evacuated to a pressure of $5\times10^{-5}$ Pa high purity (99.9999%) argon and oxygen (99.999 %) gas were introduced. The ratio of argon and oxygen is remained on 50/5.0. The target used was a high purity (99.999%) Ti plate, 60 mm in diameter and 3 mm in thickness. The output voltage to the radio frequency sputter gun was 1020 V. During sputtering, the chamber pressure was maintained at 2Pa. Concurrently, the film was heated in situ at various temperatures (100~700 $^o$C) to ensure good quality $TiO_2$ film. Then, a Co nano-layer was grown onto $TiO_2$ layer by direct current magnetron sputtering. A high purity (99.98 %) Co plate, 60 mm in diameter and 3 mm in thickness, was used as target. When the chamber was evaluated to a pressure of $5\times10^{-5}$ Pa again, only high purity (99.9999%) argon gas was introduced. The direct current sputtering power was about 26 W. And finally, another anatase $TiO_2$ was also grown by radio frequency reactive magnetron sputtering as mentioned above in order to form the $TiO_2$-Co-$TiO_2$ sandwich film. The film thickness d was determined by surface profilometry using a DEKTAK 3 Alpha-step instrument. The deposition rate was obtained by dividing d by sputtering time. In our experiment, the sputtering rate of $TiO_2$ was as large as ~2.5 nm / min, and sputtering rate of Co was about 0.28 nm / s. The sputtering system was controlled on computer.

The Co-$TiO_2$ sandwich composite films were characterized by X-ray diffraction (XRD), Raman scattering and X-ray photoemission spectra (XPS). X-ray diffraction (XRD) studies were carried out on a rotating anode D/max-rB type X-ray diffractometer operating at voltage of 40 kV and at current of 150 mA, Cu K$\alpha_1$ radiation ($\lambda$ = 0.154 nm) monochromated with a graphite sample monochromator. X-ray photoemission spectra data were obtained with a VG ESCALAB MK II spectrometer using Mg K$_\alpha$ source (1253.6 eV) radiation operating at 12 kV and 20 mA. All sample measured were calibrated with respect to the C1s peak at 284.6 eV. Raman scattering spectra were recorded by using SPEX-1403 laser Raman spectrometer with a typical resolution of 1cm$^{-1}$ in the measured frequency region. All spectra reported here were measured in backscattering geometry using 514.5 nm line Ar$^+$ laser excitation at the room temperature. The output laser power operated at 200 mW to avoid sample heating. Optical absorption spectra, neglecting reflection losses, were measured with a Perkin-Elmer Lambda 9 double beam spectrophotometer in the range of 300 to 1000 nm at room temperature. The magnetic properties were measured on an LDJ-model 9600 vibrating sample magnetometer at room temperature.

3. Results and discussion

In our experiment, the thickness of TiO$_2$ layers in the TiO$_2$-Co-TiO$_2$ film was contained a constant about 80 nm, and the thickness of Co layer was varied from 3 nm to 20 nm. Fig.1 shows X-ray diffraction patterns of the Co-TiO$_2$ sandwich film and the anatase TiO$_2$ film indicated, which were grown on glass substrate by alternately using radio frequency reactive magnetron sputtering at power of 250 W, at Ar/O$_2$ ratio of 50/5.0 and at substrate temperature of 500 $^\circ$C. For the anatase TiO$_2$, one can see seven diffracted lines signed (101), (004), (200), (211), (204), (220) and (215). The XRD data is agreeable with anatase TiO$_2$ [9,10]. The cell dimensions calculated from the diffracted lines are a = 0.378 nm and c = 0.951 nm. No diffracted lines corresponding to other polymorphs of TiO$_2$, rutile or brookite, are present. The average grain size d of TiO$_2$ estimated from the XRD profiles using Scherrer's equation [11] d

$= k\lambda / \beta \cos\theta$, where k is a constant (shape factor, about 0.9), λ is the x-ray wavelength, β is the full-width at half-maximum (FWHM) of the diffraction line and θ is the diffraction angle. In $TiO_2$ film, the mean particle size of 13.2 nm is estimated. For $Co-TiO_2$ sandwich film, one can see some feature diffraction peaks of anatase $TiO_2$ and a Co (111) appeared. The XRD peak signals were weak because the film was very thin (about 150 nm). The XRD pattern of $Co-TiO_2$ sandwich film revealed that the film is a mixture of the anatase $TiO_2$ and the cubic cobalt.

Fig.2 indicates Raman scattering spectra of the anatase $TiO_2$ film and the $Co-TiO_2$ sandwich film. One can see that the four Raman scattering peaks of anatase $TiO_2$ are at approximately 144 $cm^{-1}$, 400 $cm^{-1}$, 515 $cm^{-1}$ and 640 $cm^{-1}$, which are assigned two $B_{1g}$ modes and two $E_g$ modes respectively. These results are agreeable with Ref. [10, 12]. In addition, we found another Raman peak at around 1106 $cm^{-1}$, which was attributed to the vibration mode of $Ti-O-SiO_2$ [13], indicating that there is a strong interaction at the interface between the $TiO_2$ film and glass substrate.

Fig. 3 shows XPS Ti 2p core level of (a) anatase $TiO_2$ and (b) $Co-TiO_2$ sandwich film. One sees that the core levels of Ti $2p_{1/2}$ and Ti $2p_{3/2}$ are at approximately 464.1 and 458.4 eV, respectively, which are assigned to the $Ti^{4+}$ ($TiO_2$) in reasonable agreement with the literature [14], with a peak separation of 5.7 eV between those two peaks. Compared with Ref. [15,16], the core levels of Ti $2p_{1/2}$ and Ti $2p_{3/2}$ have a small shift of 0.1 eV.

Fig. 4 shows XPS O 1s core level of (a) anatase $TiO_2$ and (b) $Co-TiO_2$ sandwich film. The core level of O 1s is at line of 529.7 eV from Fig. 4, which is attributed to titanium dioxide at the surface. This data is approximately agreeable with the main peak at 529.9 eV of

TiO$_2$ [15].

Fig. 5 shows XPS Co 2p core level of Co-TiO$_2$ sandwich film. The core level of Co 2p$_{1/2}$ and Co 2p$_{3/2}$ are at 797.1 and 782.2 eV respectively, with a peak separation of 14.9 eV between those two peaks. Compared Ref. [15], it shows that the Co is metal state in the anatase TiO$_2$. However, the core levels of Co 2p$_{3/2}$ and Co 2p$_{1/2}$ in anatase TiO$_2$ have a chemical shift of 0.42 eV and 0.43 eV, respectively. This demonstrates that there is an interaction between Co and TiO$_2$.

Fig. 6 indicates optical absorption of Co-TiO$_2$ sandwich films with the thickness of Co nano-layer indicated. For nanocrystalline anatase TiO$_2$ film (the thickness of Co nano-layer is zero), the optical absorption edge blue shifted to 320.9 nm (3.86 eV) compared to that of bulk anatase TiO$_2$ (3.20 eV) [17], which was originated to the quantum confinement effect. However, when the Co nano-layer inside the two-anatase TiO$_2$ layers, Co atoms will self-diffuse to the matrix of TiO$_2$ and change the band gap of Co-TiO$_2$ films. From Fig. 6, one sees that the optical absorption of Co-TiO$_2$ sandwich composite films red shift to 321.4, 330.8, 333.5 and 345.2 nm, which corresponding to the thickness of the Co nano-layer of 3, 9, 12 and 48 nm, respectively. This indicates that the band gap of anatase TiO$_2$ is widened with increasing the Co nano-layer. The variation of band gap with the thickness d of the Co nano-layer in the Co-TiO$_2$ sandwich composite films is presented in Fig. 7. It shows that the threshold α of fundamental absorption of the Co-TiO$_2$ sandwich film may be described by the expression:

$$\alpha = A(E - E_g)^m \quad (1)$$

Where E is the optical band gap of the Co-TiO$_2$ sandwich film, E$_g$ is the optical band gap of pure anatase $TiO_2$, and A is a constant. The value of m may be taken m = 2, a characteristic value for the indirect allowed transition dominates over the optical absorption [18].

Fig. 8 indicates optical absorption spectra of Co-$TiO_2$ sandwich film deposited at the power of 250 W, at the $Ar/O_2$ ratio of 50/5.0 and the different substrate temperatures indicated. It shows that the optical absorption edges have a red shift with increasing substrate temperatures. The results of these optical absorption edges are about 339.7, 349.8 and 400.6 nm, which corresponding to the substrate temperature of 300 $^oC$, 350 $^oC$ and 400 $^oC$ respectively. These can be interpreted that the band gaps of Co-$TiO_2$ film decrease with much of Co atoms diffusing to the matrix of $TiO_2$, which attributed to the increasing substrate temperature.

Fig. 9 presents hysteresis loop for Co-$TiO_2$ sandwich film in magnetic field parallel to the film plane. This film has two 80 nm $TiO_2$ layers and 12 nm Co layer. It shows that the saturation magnetization, remanent magnetic induction and coercivity are about 4.35 emu/g, 1.02 emu/g and 569.6 Oe respectively. This demonstrates that the Co-$TiO_2$ sandwich film has ferromagnetic properties.

4. Conclusions

In summary, the Co-$TiO_2$ sandwich composite films have been recently grown on glass and silicon substrates by alternately using radio frequency reactive and direct current magnetron sputtering. The microstructural properties of these films were characterized with Raman spectra and X-ray photoemission spectra (XPS). It shows an anatase $TiO_2$ containing Co nano-layer. Optical absorption spectra have red shift with Co nano-layer increasing, which originated from the quantum tunnel and quantum confinement effects. Magnetic properties show that the saturation magnetization, remanent magnetic induction and coercivity are about 4.35 emu/g, 1.02 emu/g and 569.6 Oe respectively.

Acknowledgements

This paper was supported by the Postdoctoral Foundation of China, and partly supported by National Natural Science Foundation of China (No.59982002) and the foundation of State Key Laboratory for Advanced Metals and Materials, University of Science and Technology Beijing.

Caption for Figures

Fig. 1 X-ray diffraction patterns of the Co-TiO$_2$ sandwich film and the anatase TiO$_2$ film

Fig.2 Raman scattering spectra of the anatase TiO$_2$ film and the Co-TiO$_2$ sandwich film

Fig. 3 XPS Ti 2p core level of (a) anatase TiO$_2$, (b) Co-TiO$_2$ sandwich film

Fig. 4 XPS O 1s core level of (a) anatase TiO$_2$, (b) Co-TiO$_2$ sandwich film

Fig. 5 XPS Co 2p core level of Co-TiO$_2$ sandwich film

Fig. 6 Optical absorption of Co-TiO$_2$ sandwich films with the thickness of Co nano-layer indicated.

Fig. 7 Variation of band gap Vs the thickness of Co-nanolayer in the Co-TiO$_2$ sandwich composite films

Fig. 8 Optical absorption spectra of Co-TiO$_2$ sandwich film deposited at different substrate temperatures indicated.

Fig. 9 Hysteresis loops for Co-TiO$_2$ sandwich film in magnetic field parallel to the film plane.

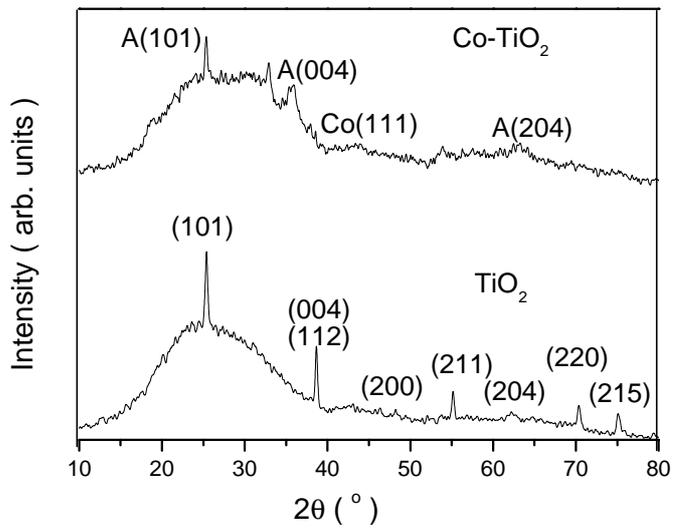

Fig. 1　X-ray diffraction patterns of the Co-TiO$_2$ sandwich film and the anatase TiO$_2$ film

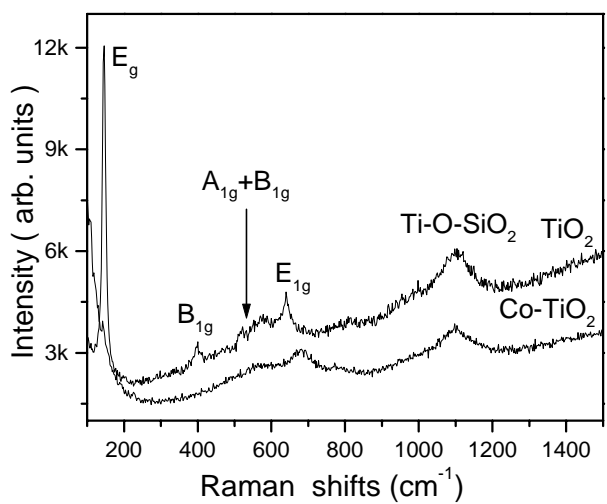

Fig.2　Raman scattering spectra of the anatase TiO$_2$ film and the Co-TiO$_2$ sandwich film

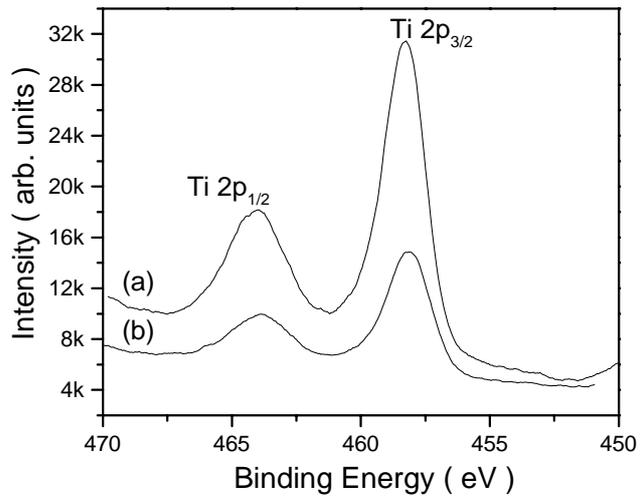

Fig. 3 XPS Ti 2p core level of (a) anatase $TiO_2$, (b) Co-$TiO_2$ sandwich film

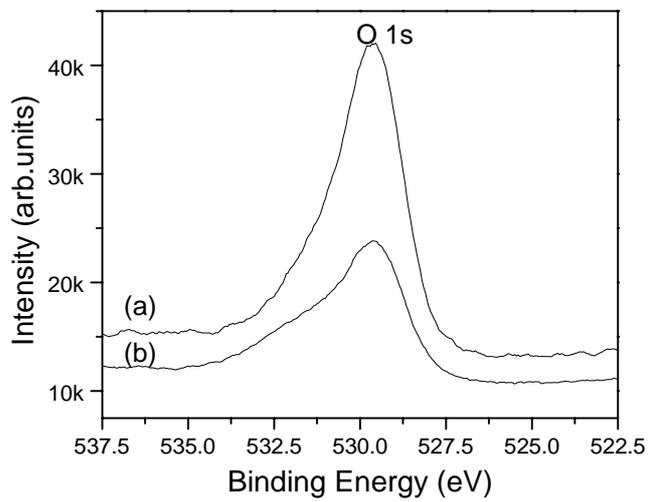

Fig. 3 XPS O 1s core level of (a) anatase $TiO_2$, (b) Co-$TiO_2$ sandwich film

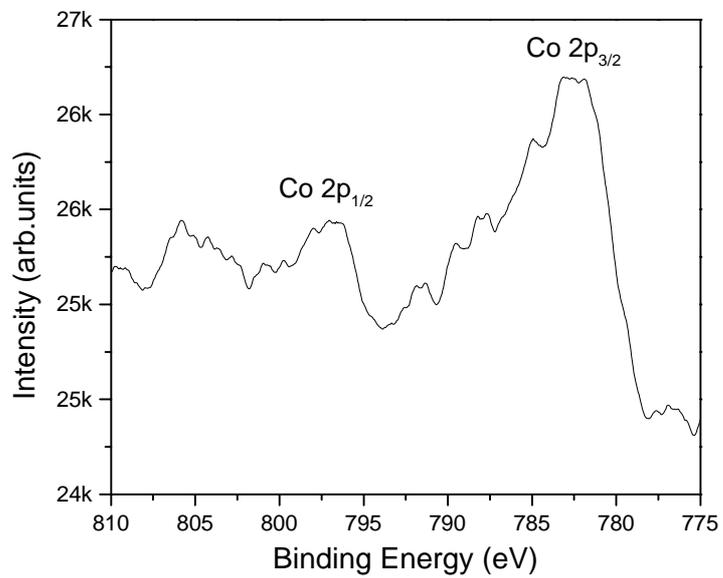

Fig. 5  XPS Co 2p core level of Co-TiO$_2$ sandwich film

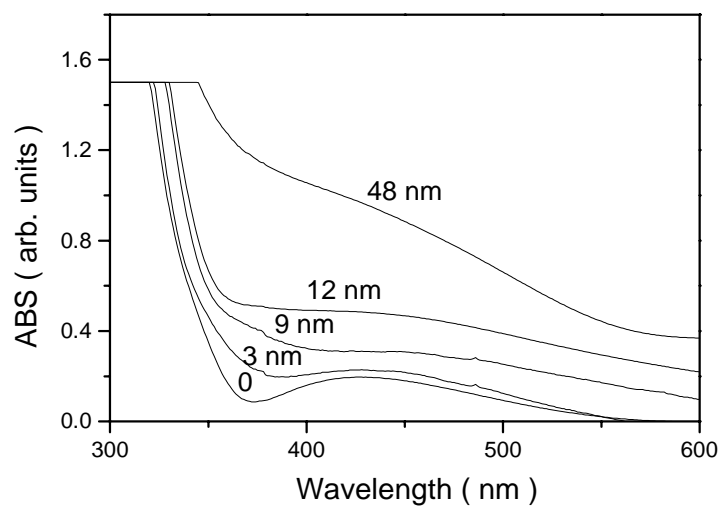

Fig. 6 Optical absorption of Co-TiO$_2$ sandwich films with the thickness of Co nano-layer indicated.

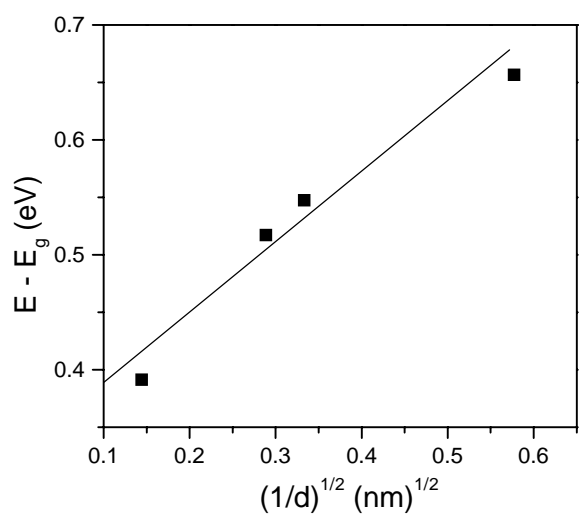

Fig. 7 Variation of band gap Vs the thickness of Co-nanolayer in the Co-TiO$_2$ sandwich composite films

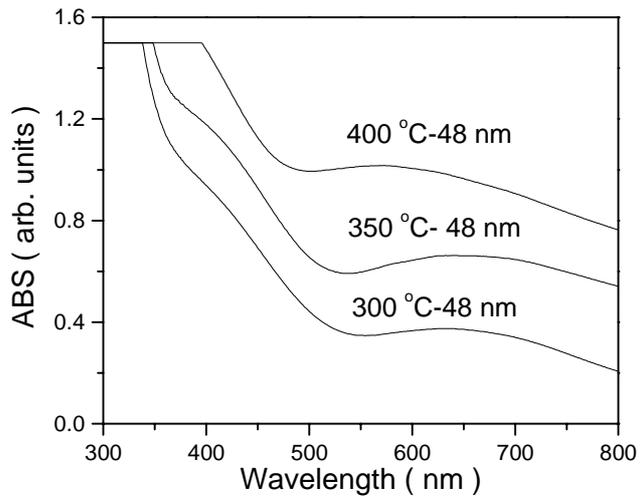

Fig. 8 Optical absorption spectra of Co-TiO$_2$ sandwich film deposited at different substrate temperatures indicated

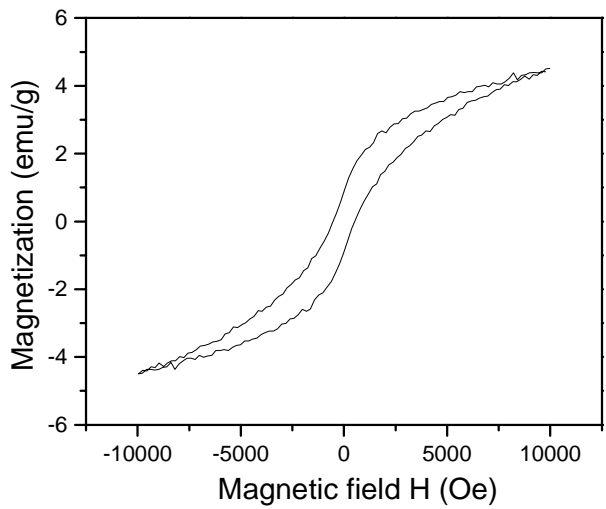

Fig. 9  Hysteresis loop for Co-TiO$_2$ sandwich film in magnetic field parallel to the film plane. The TiO$_2$ and Co layers are about 80 nm and 12 nm, respectively.